\documentclass[mathleft
]{an}
\usepackage{graphicx}
\usepackage{times}
\newcommand{\aap}{A\&A }
\newcommand{\aaps}{A\&AS }
\newcommand{\skytel}{S\&T}

\begin{document}

\Pagespan{789}{}
\Yearpublication{2006}%
\Yearsubmission{2005}%
\Month{11}%
\Volume{999}%
\Issue{88}%

\title{Planetary transit observations at the University Observatory Jena: TrES-2\thanks{Based on observations obtained with telescopes of the University Observatory Jena, which is operated by the Astrophysical Institute of the Friedrich-Schiller-University Jena and the 80cm telescope of the Wendelstein Observatory of the Ludwig-Maximilians-University Munich.}}

\author{St. Raetz\inst{1}\fnmsep\thanks{Corresponding author:
  \email{straetz@astro.uni-jena.de}\newline}
\and M. Mugrauer\inst{1}
\and T. O. B. Schmidt\inst{1}
\and T. Roell\inst{1}
\and T. Eisenbeiss\inst{1}
\and M. M. Hohle\inst{1,4}
\and A. Koeltzsch\inst{1}
\and M. Va{\v n}ko\inst{1}
\and Ch. Ginski\inst{1}
\and C. Marka\inst{1}
\and M. Moualla\inst{1}
\and N. Tetzlaff\inst{1}
\and A. Seifahrt\inst{1,2}
\and Ch. Broeg\inst{3}
\and \\ J. Koppenhoefer\inst{5}
\and M. Raetz\inst{6}
\and R. Neuh{\"a}user\inst{1}
}
\titlerunning{Transit observation at the University Observatory Jena}

\institute{
Astrophysikalisches Institut und Universit{\"a}ts-Sternwarte Jena, Schillerg{\"a}{\ss}chen 2-3, 07745 Jena, Germany
\and 
Institut f{\"u}r Astrophysik, Georg-August-Universit{\"a}t, Friedrich-Hund-Platz 1, 37077 G{\"o}ttingen, Germany
\and 
Space Research and Planetary Sciences, Physikalisches Institut, University of Bern, Sidlerstra{\ss}e 5, 3012 Bern, Switzerland
\and
Max Planck Institute for Extraterrestrial Physics, Giessenbachstra{\ss}e, 85748 Garching, Germany
\and
University Observatory Munich, Scheinerstrasse 1, 81679 M\"unchen, Germany
\and
Private observatory Raetz, Stiller Berg 6, 98587 Herges-Hallenberg, Germany}

\received{2008 Dec 9}
\accepted{2009 Apr 3}
\publonline{2009 May 30}


\abstract{We report on observations of several transit events of the transiting planet TrES-2 obtained with the Cassegrain-Telskop-Kamera at the University Observatory Jena. Between March 2007 and November 2008 ten different transits and almost a complete orbital period were observed. Overall, in 40 nights of observation 4291 exposures (in total 71.52\,h of observation) of the TrES-2 parent star were taken. With the transit timings for TrES-2 from the 34 events published by the TrES-network, the Transit Light Curve project and the Exoplanet Transit Database plus our own ten transits, we find that the orbital period is $P=(2.470614\pm 0.000001)\,$d, a slight change by $\sim$\,0.6\,s compared to the previously published period. We present new ephemeris for this transiting planet.\\ Furthermore, we found a second dip after the transit which could either be due to a blended variable star or occultation of a second star or even an additional object in the system.\\ Our observations will be useful for future investigations of timing variations caused by additional perturbing planets and/or stellar spots and/or moons.} 

\maketitle

\section{Introduction}

TrES-2 is the second transiting hot Jupiter discovered by the Trans-atlantic Exoplanet Survey (TrES, O'Donovan et al. 2006). The planet orbits the nearby 11th magnitude G0\,V main-sequence dwarf GSC\,03549-02811 every $\sim$2.5 days. From high-resolution spectra, Sozzetti et al. (2007) derived accurate values of the stellar atmospheric parameters of the TrES-2 parent star such as effective temperature, surface gravity and metallicity. With the help of detailed analysis of high precision z-band photometry and light curve modeling of the 1.4\% deep transit by the Transit Light Curve (TLC) Project (Holman et al. 2007), estimates of the planetary parameters could also be determined.\\ One goal of the TLC project is to measure variations in the transit times and light curve shapes that would be caused by the influence of additional bodies in the system (Agol et al. 2005; Holman \& Murray 2005; Steffen \& Agol 2005). The large impact parameter of TrES-2 makes the duration more sensitive to any changes. Hence, TrES-2 is an excellent target for the detection of long-term changes in transit characteristics induced by orbital precession (Miralda-Escud\'e 2002). \\ We have started high precision photometric observations at the University Observatory Jena in fall 2006. In this work we use the transit method to observe the transiting planet TrES-2. We paid special attention to the accurate determination of transit times in order to identify precise transit timing variations that would be indicative of perturbations from additional bodies and to refine the orbital parameters of the system. First results were presented in Raetz et al. (2009a). \\ TrES-2 lies within the field of view of the NASA \textit{Kepler} mission (Borucki et al 2003; Basri et al 2005). During the four year mission, \textit{Kepler} will observe nearly 600 transits of TrES-2 (O'Donovan et al. 2006). The precision of \textit{Kepler} will be extremely sensitive to search  for additional planets in the TrES-2 system through their dynamical perturbations. \\ In this paper, we describe the observations, the data reduction and the analysis procedures. Furthermore, we present results for TrES-2 that we obtained from our observations at the University Observatory Jena.

\section{Observations}
\label{observations}

\subsection{University Observatory Jena Photometry}

\begin{figure}[h]
  \centering
  \includegraphics[width=0.48\textwidth]{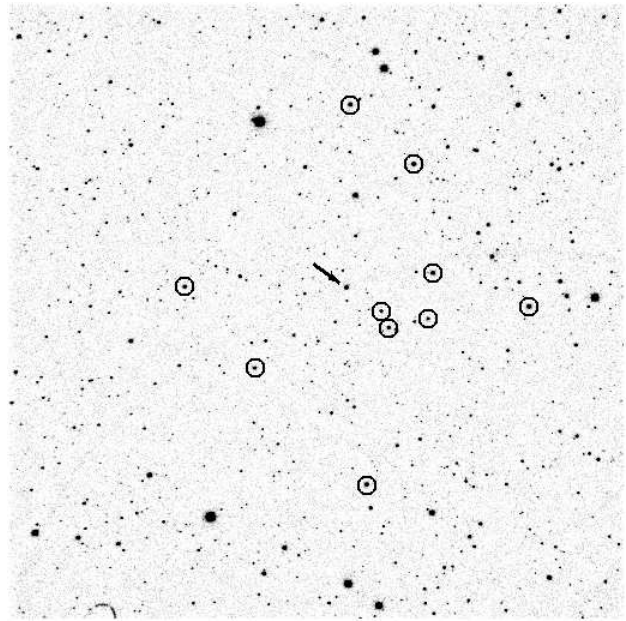}
  \caption{The 37.7\,$'\times$ 37.7\,$'$ field of view of the CTK ($I$-band, exposure time 60\,s). The arrow marks the TrES-2 host star, the circles the 10 most constant comparison stars. North is up; east is to the left.}
  \label{FoV_TrES2}
\end{figure}

Most observations were carried out at the University Observatory Jena which is located close to the village Gro{\ss}- schwabhausen, 10\,km west of the city of Jena.\\ Mugrauer (2009) describes the instrumentation and operation of the system. Our transit observations are carried out with the CTK (\textit{\underline{C}assegrain \underline{T}eleskop \underline{K}amera), the CCD imager operated at the 25\,cm auxiliary telescope of the University Observatory Jena.} \\ The CTK CCD-detector consists of 1024\,$\times$\,1024 pixel with a pixel scale of about 2.2\,$''$/pixel which yields a field of view of 37.7\,$'\,\times\,$37.7\,$'$ (for more details see Mugrauer 2009). \\ We started our observations in November 2006. Between March 2007 and November 2008, we used 49 clear nights for our transit observations. Part of the time we observed known transiting planets in  Bessell $R$ and $I$ band.\\ For our TrES-2 observations, started in March 2007, we used 43 nights from March 2007 to November 2008. Due to the weather conditions 3 nights were not useable. We observed nine different transits. These transits correspond to epochs 87, 108, 138, 163, 165, 174, 278, 316 and 318 of the ephemeris given by Holman et al. (2007):
\begin{eqnarray}
\label{Ephemeris_Holman}
\begin{array}{r@{.}lcr@{.}l}
T_{\mathrm{c}}(E)=(2453957 & 63479 & + & E\cdot 2 & 470621)\,\mathrm{d}\\
\pm0 & 00038 &  & \pm0 & 000017
\end{array}
\end{eqnarray}
To get a phase diagram of one full orbital period, we observed TrES-2 even in times out of transit. All TrES-2 observations were taken in $I$-band with 60 s exposure time. We achieve a mean cadence of the data points of 1.4\,min (readout time of the CTK around 24\,s). The mean photometric precision of the $V\,\approx$\,11\,mag bright TrES-2 host star is 0.007 mag. Fig. \ref{FoV_TrES2} shows the field of view of the CTK around the TrES-2 host star.
\begin{figure*}[tH]
  \centering
  \includegraphics[width=0.9\textwidth]{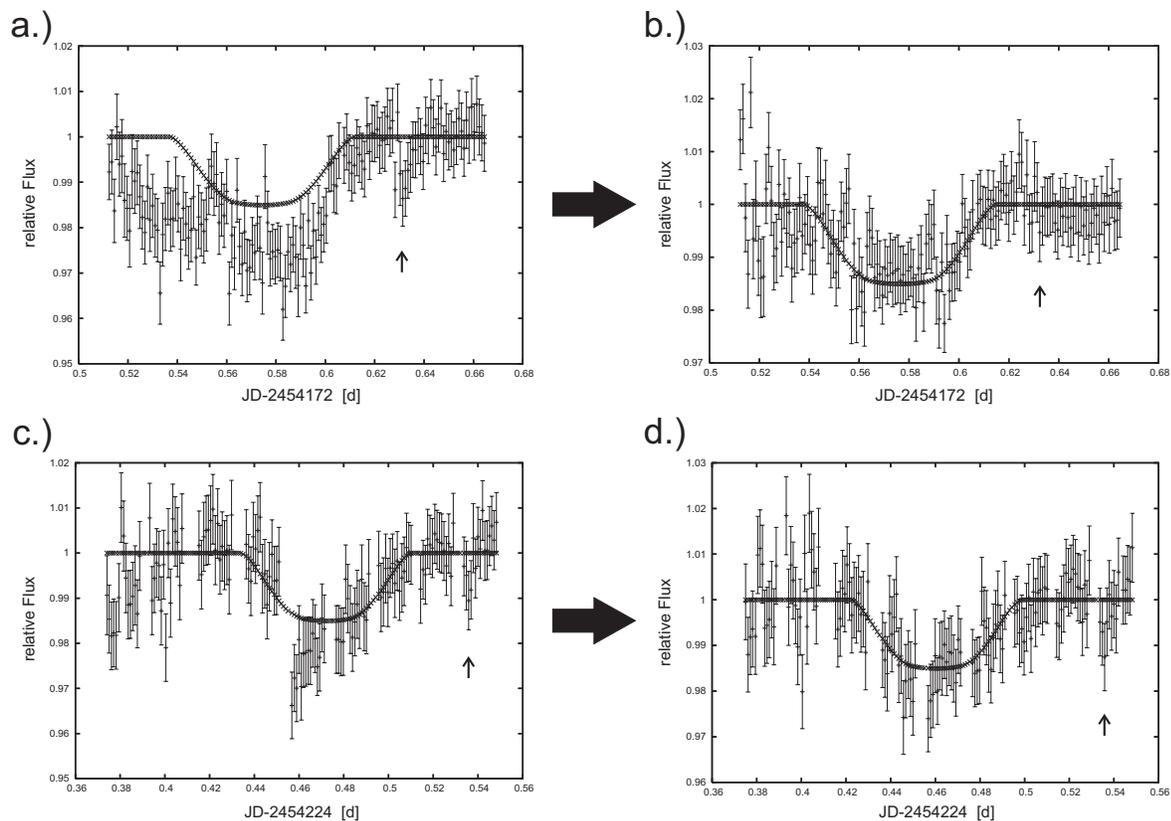}
  \caption{Two different transits of TrES-2 before and after using Sys-Rem. \textbf{a.)} The transit from 2007 March 13 before Sys-Rem \textbf{b.)} The transit from 2007 March 13 after Sys-Rem \textbf{c.)} The transit from 2007 May 3 before Sys-Rem \textbf{d.)} The transit from 2007 May 3 after Sys-Rem. The arrow marks the time of the second dip discussed in Section \ref{2nd_dip}}
  \label{TrES2_Sysrem}
\end{figure*}
\begin{figure*}[tH]
  \centering
  \includegraphics[width=0.9\textwidth]{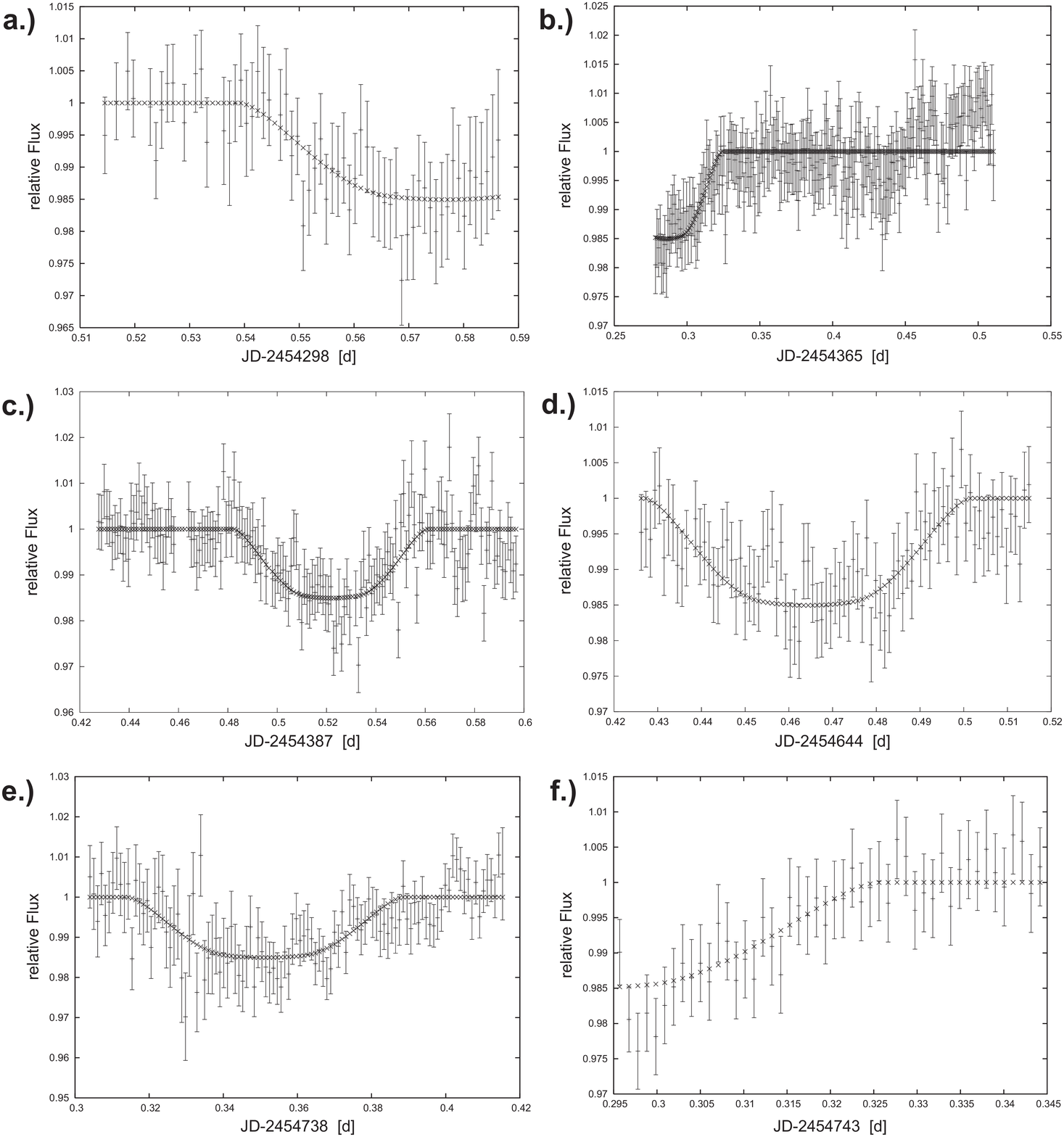}
  \caption{$I$ band photometry of six TrES-2-transits observed at the University Observatory Jena: \textbf{a.)} 2007 July 17 \textbf{b.)} 2007 September 21 \textbf{c.)} 2007 October 13 \textbf{d.)} 2008 June 26 \textbf{e.)} 2008 September 28 \textbf{f.)} 2008 October 3}
  \label{Lightcurves_TrES2_Gsh}
\end{figure*}

\subsection{Additional Photometry}

In 2007 July and September, we observed two transit events at the Wendelstein Observatory of the university of Munich. The transits from 2007 July 26 and 2007 September 16 include 157 and 137 $R$-band 30\,s exposures, respectively. According to the ephemeris provided by Holman et al. (2007), these transits correspond to epochs 142 and 163. All 294 images were acquired using the MONICA (MOno-chromatic Image CAmera, Roth 1990) CCD camera, at the Cassegrain focus of the 0.8\,m Wendelstein telescope, with a pixel scale of 0.5 $''$/pixel and a field of view of 8.5$'\,\times$\,8.5$'$. \\ The pre-reduction was done using standard reduction software (MUPIPE\footnote{http://www.usm.lmu.de/$\sim$arri/mupipe/}) specifically developed at the Munich Observatory for the MONICA CCD camera (G{\"o}ssl \& Riffeser 2002). For these observations the mean photometric precision is 0.005\,mag.\\ Because of the good weather conditions on 2007 September 16 we recruited an amateur astronomer (M.R.) to observe this transit of TrES-2. As Naeye (2004) and McCullough et al. (2006) showed, amateur astronomers can produce photometry with sufficient quality to detect a transit. With an 8-inch ($\sim$\,20\,cm) Schmidt-Cassegrain telescope (f/D\,=\,10) and a ST-6 CCD camera (field of view 13.6\,$'\,\times\,$10.2\,$'$) without filter (white-light observations) located in Herges-Hal- lenberg, Germany, we reached a photometric precision of 0.01\,mag. With this observation we have three independent measurements of the transit from 2007 September 16 (epoch 163 according to eq. \ref{Ephemeris_Holman}), see Fig. \ref{TrES2_07_09_16}.

\section{Data Reduction and analysis of time series}

\subsection{In general}
\label{general}

We calibrate the images of our target field using the standard IRAF\footnote{IRAF is distributed by the National Optical Astronomy Observatories, which are operated by the Association of Universities for Research in Astronomy, Inc., under cooperative agreement with the National Science Foundation.} procedures \textit{darkcombine}, \textit{flatcombine} and \textit{ccdproc}. We did not correct for bad pixel (see Raetz et al. 2009b). \\ First we perform aperture photometry by using the IRAF task \textit{chphot} (see Raetz et al. 2009b).\\ For differential photometry we use an algorithm that calculates an artificial comparison star by taking the weighted average of a maximum number of stars (all available field stars). To compute the best possible artificial comparison star we successively sort out all stars with low weights (stars that are not on every image, stars with low S/N and variable stars). As result, we get an artificial comparison star made of the most constant stars with the best S/N in the field (Broeg et al. 2005, Raetz et al. 2009b).\\ In the third step we correct for systematic effects by using the Sys-Rem detrending algorithm which was developed by Tamuz et al. (2005) and implemented by Johannes Koppenhoefer, see Fig. \ref{TrES2_Sysrem}. We showed in Raetz et al. (2009b) that the usage of Sys-Rem is not possible in every case of observations at the University Observatory Jena.

\subsection{The case of TrES-2}

\begin{figure}[h]
  \centering
  \includegraphics[width=0.48\textwidth]{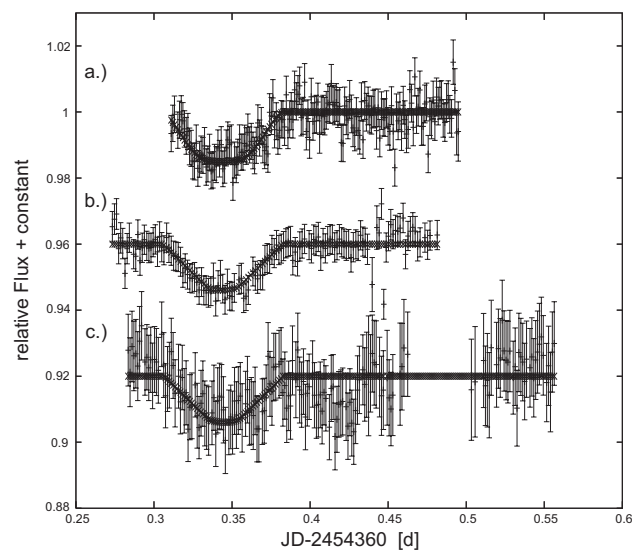}
  \caption{The TrES-2-transit from 2007 September 16 with three different telescopes: \textbf{a.)} $I$-band photometry from Jena \textbf{b.)} Wendelstein $R$-band observations \textbf{c.)} White light amateur observations}
  \label{TrES2_07_09_16}
\end{figure}

\begin{figure}[h]
  \centering
  \includegraphics[width=0.4\textwidth]{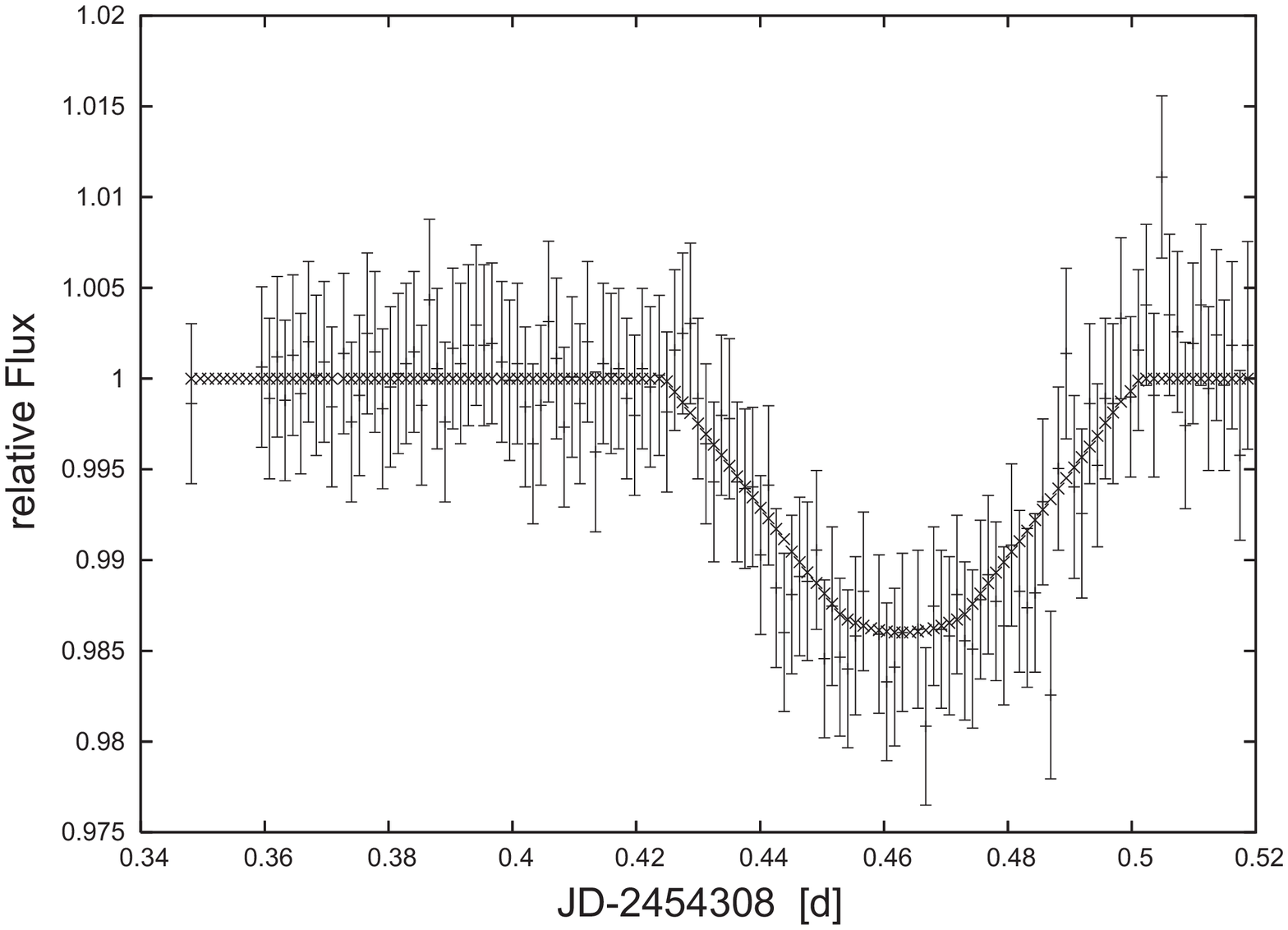}
  \caption{Wendelstein $R$-band photometry from 2007 July 26.}
  \label{TrES2_07_07_26_W}
\end{figure}

After calibrating all images, we perform aperture photometry on all available field stars. We found 1294 stars - the TrES-2 host star and 1293 comparison stars - in the CTK field of view. We used an aperture of 5 pixels (11.03$''$) radius and an annulus for sky subtraction ranging in radius from 15 to 20 pixels, centered on each star.\\ To get the best possible result for the transit light curves, we reject all comparison stars that could not be measured on every image, faint stars with low S/N and variable stars which could introduce disturbing signals to the data. We have done this analysis individually for every night. To get comparable values for every night, we search for those comparison stars that belong always to the most constant stars. Finally we use for every night the same 10 constant comparison stars to calculate the artificial comparison stars (see Fig. \ref{FoV_TrES2}). With this method we could improve the preliminary results given in Raetz et al. (2009a).\\ For the Wendelstein observations we chose an aperture size of 10 pixels (5$''$). And for sky subtraction we use a ring-shaped annulus having an inner/outer radius of 15/20 pixels, respectively. We were not able to use the same 10 comparison stas as in the case of observations at the University Observatory Jena because they do not lie within the 8.5$'\,\times$\,8.5$'$ field of view of MONICA. From 225 objects in this field we could calculate the artificial comparison star for both nights using the seven most constant stars. \\ For the amateur observations only 147 stars could be measured within an aperture of 5 pixels (11.8$''$) and an annulus for sky substraction similar to observations at the University Observatory Jena and on Wendelstein. Again it was not possible to use the 10 best comparison stars froUniversity Observatory Jena observations or the seven comparison stars from Wendelstein observations. In this case the artificial comparison star consists of the 15 most contant stars in the ST-6 field.\\ Finally the TrES-2 host star is compared to the artificial comparison star to get the differential magnitudes for every image.\\ In the resulting light curves of the TrES-2 host star and the comparison stars we used Sys-Rem. The algorithm works without any prior knowledge of the effects. The number of effects that should be removed from the light curves is selectable and can be set as a parameter. As mentioned in section \ref{general} it was not possible to use Sys-Rem for every night. In Table \ref{our_transits} we summarize how many effects we removed from the light curves by using Sys-Rem before the transit itself is identified as systematic effect. Zero means that we could not apply Sys-Rem. Fig. \ref{TrES2_Sysrem} shows for example the effect of Sys-Rem for two different transits of TrES-2.\\ The resulting light curves from the University Observatory Jena observations can be found in Fig. \ref{Lightcurves_TrES2_Gsh} and \ref{TrES2_07_09_16}a, the Wendelstein observations in Fig. \ref{TrES2_07_09_16}b and \ref{TrES2_07_07_26_W} and the amateur observations in Fig. \ref{TrES2_07_09_16}c.
\begin{table}
\caption{Information of observations and analysis for all measured transit events}
\label{our_transits}
\begin{tabular}{lcccc}\hline
Date & Observatory & Transit & Points$^{a}$ & removed \\
& & & & Effects$^{b}$ \\
\hline
2007 Mar 13 & Jena & Full & 96 & 1 \\
2007 May 3  & Jena & Full & 75 & 1 \\
2007 Jul 17 & Jena  & Partial & 52 & 0 \\
2007 Jul 26 & Wendelstein & Full & 76 & 0 \\
2007 Sep 16 & Jena  & Full & 78 & 1 \\
2007 Sep 16 & Wendelstein & Full & 65 & 2 \\
2007 Sep 16 & Amateur & Full & 75 & 1 \\
2007 Sep 21 & Jena & Partial & 55 & 0 \\
2007 Oct 13 & Jena & Full & 89 & 1 \\
2008 Jun 26 & Jena & Full & 85 & 1 \\
2008 Sep 28 & Jena & Full & 96 & 1 \\
2008 Oct 03 & Jena & Partial & 43 & 0 \\
\hline
\end{tabular}
\\
$^{a}$ Number of data points covering the transit; i.e., within the phases 0.98-1.02.\\
$^{b}$ Number of effects that are removed from the light curves by using Sys-Rem\\
\end{table}

\begin{table}
\caption{Summary of all previously known and new transit times of TrES-2.}
\label{Transit_times_TrES2}
\begin{tabular}{llr@{\,$\pm$\,}l}
\hline
Observer & Epoch$^{a}$ & \multicolumn{2}{c}{HJD (Midtransit)} \\ \hline
TrES-Network$^{b}$ & 0 & 2453957.63580 & 0.00100 \\
ETD$^{c}$ & 12 & 2453987.28000 & 0.00800 \\
TLC-Project$^{d}$ & 13 & 2453989.75286 & 0.00029 \\
& 15 & 2453994.69393 & 0.00031 \\
ETD & 19 & 2454004.57500 & 0.00140 \\
& 25 & 2454019.40150 & 0.00600 \\
TLC-Project & 34 & 2454041.63579 & 0.00030 \\
This work & 87 & 2454172.57670 & 0.00160 \\
ETD & 106 & 2454219.52050 & 0.00600 \\
This work & 108 & 2454224.46176 & 0.00250 \\
ETD & 127 & 2454271.39911 & 0.00297\\
& 130 & 2454278.81790 & 0.00600 \\
This work & 138 & 2454298.57880 & 0.00240 \\
& 142 & 2454308.46448 & 0.00130 \\
ETD & 142 & 2454308.46240 & 0.00600 \\
& 142 & 2454308.46300 & 0.00180 \\
& 151 & 2454330.70130 & 0.00200 \\
& 155 & 2454340.58350 & 0.00120 \\
& 157 & 2454345.51390 & 0.00160 \\
& 157 & 2454345.51990 & 0.00120 \\
& 157 & 2454345.52350 & 0.00150 \\
This work & 163$^{e}$ & 2454360.34550 & 0.00109 \\
& 165 & 2454365.28746 & 0.00210 \\
ETD & 170 & 2454377.63810 & 0.00070 \\
& 170 & 2454377.64230 & 0.00120 \\
This work & 174 & 2454387.52220 & 0.00150 \\
ETD & 229 & 2454523.40970 & 0.00080 \\
& 242 & 2454555.52621 & 0.00123 \\
& 242 & 2454555.52360 & 0.00090 \\
& 259 & 2454597.52250 & 0.00120 \\
& 268 & 2454619.75990 & 0.00130 \\
& 272 & 2454629.64510 & 0.00240 \\
This work & 278 & 2454644.46608 & 0.00140 \\
ETD & 278 & 2454644.46440 & 0.00180 \\
& 280 & 2454649.41490 & 0.00330 \\
& 281 & 2454651.87560 & 0.00070 \\
& 293 & 2454681.52240 & 0.00210 \\
& 304 & 2454708.69870 & 0.00110 \\
& 310 & 2454723.51790 & 0.00190 \\
This work & 316 & 2454738.35215 & 0.00200 \\
ETD & 316 & 2454738.35045 & 0.00090 \\
This work & 318 & 2454743.28972 & 0.00180 \\
ETD & 321 & 2454750.70010 & 0.00110 \\
& 333 & 2454780.34690 & 0.00220 \\
\hline
\end{tabular}
\\ $^{a}$ according to the ephemeris of Holman et al. (2007)\\
$^{b}$ from O'Donovan et al. (2006)\\
$^{c}$\,from various observers collected in Exoplanet Transit Database, http://var.astro.cz/ETD \\
$^{d}$ from Holman et al. (2007)\\
$^{e}$ weighted average of the values of all three observatories (see Table \ref{our_transits})
\end{table}


\section{Determination of the midtransit time}

Because the output of the IRAF task \textit{phot} are magnitudes we convert our data to relative flux. We then calculate a weighted average of the pre-ingress and post-egress data and divided all data points by this value, in order to normalize the flux.\\ To determine the time of the center of the transit we fit an analytic light curve on the observed light curve. This analytic light curve was calculated with the stellar and planetary parameters given by Holman et al. (2007) and Sozzetti et al. (2007). To get the best fit we compare the analytic light curve with the observed light curve until the $\chi^{2}$ is minimal. The best fits of the analytic light curve to our observed data are shown as crosses in Fig. \ref{TrES2_Sysrem}, \ref{Lightcurves_TrES2_Gsh}, \ref{TrES2_07_09_16} and \ref{TrES2_07_07_26_W}. With the help of the $\chi^{2}$-test we could determine the time of the midtransit even in the case of partial transits. We summarize the observed midtransit times in Table \ref{Transit_times_TrES2} marked with ''This work''. We give the 1-$\sigma$ error bars.

\section{Ephemeris}

\begin{table*}
\centering
\caption{$UBVRI$ photometry of two Landolt standard stars and four secondary standard stars (from Galad\'i-Enr\'iquez et al. 2000)}
\label{UBVRI_photometry}
\begin{tabular}{lccr@{\,$\pm$\,}lr@{\,$\pm$\,}lr@{\,$\pm$\,}lr@{\,$\pm$\,}lr@{\,$\pm$\,}l}
\hline
star & $\alpha_{2000}$ & $\delta_{2000}$ & \multicolumn{2}{c}{$V$} & \multicolumn{2}{c}{$B\,-\,V$} & \multicolumn{2}{c}{$U\,-\,B$} & \multicolumn{2}{c}{$V\,-\,R$} & \multicolumn{2}{c}{$V\,-\,I$} \\ \hline
SA\,44\,28$^{a}$ & 00 29 04.00 & +30 23 12.0 & 11.329 & 0.002 & 0.726 & 0.001 & 0.200 & 0.004 & 0.394 & 0.001 & 0.764 & 0.002 \\
SA\,44\,113$^{a}$ & 00 29 38.00 & +30 23 18.0 & 11.713 & 0.006 & 1.206 & 0.002 & 0.996 & 0.019 & 0.667 & 0.003 & 1.229 & 0.005 \\
8$^{b}$ & 00 29 06.45 & +30 25 29.6 & 14.539 & 0.006 & 0.765 & 0.005 & 0.276 & 0.006 & 0.426 & 0.004 & 0.809 & 0.007 \\
9$^{b}$ & 00 29 07.88 & +30 18 07.1 & 13.099 & 0.007 & 0.609 & 0.003 & -0.080 & 0.026 & 0.355 & 0.006 & 0.712 & 0.004 \\
22$^{b}$ & 00 29 20.55 & +30 26 08.0 & 10.187 & 0.003 & 1.118 & 0.004 & 0.971 & 0.023 & 0.583 & 0.009 & 1.102 & 0.005 \\
29$^{b}$ & 00 29 33.25 & +30 24 05.6 & 13.563 & 0.008 & 0.633 & 0.006 & -0.043 & 0.021 & 0.348 & 0.006 & 0.666 & 0.005 \\
\hline
\end{tabular}
\\
\begin{flushleft}
$^{a}$ Landolt (1983)\\
$^{b}$ Numbers of standard stars in field \#1 defined by Galad\'i-Enr\'iquez et al. (2000)\\
\end{flushleft}
\end{table*}

\begin{figure}[h]
  \centering
  \includegraphics[width=0.48\textwidth]{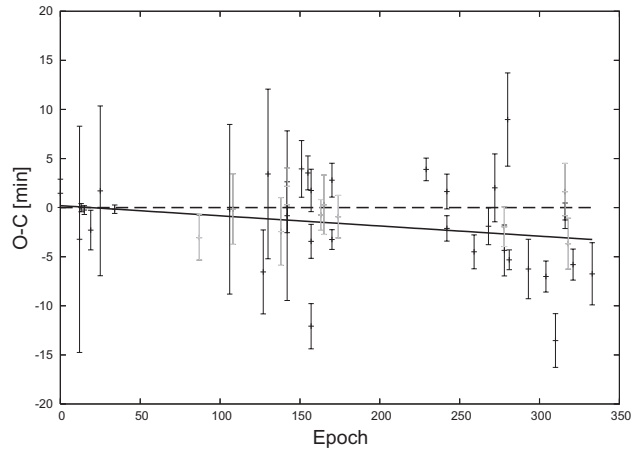}
  \caption{Transit timing residuals for TrES-2. The gray data points correspond to the measurements from the University Observatory Jena. The dashed line shows the ephemeris given by Holman et al. (2007). The best-fitting line (solid line) is plotted, representing the updated ephemeris given in equation \ref{ephemeris_new_TrES2}.}
  \label{O_C_TrES2}
\end{figure}

In addition to the transits observed by us we could find 34 transit times from 2006-2008 in the literature. The altogether 44 transits are summarized in Table \ref{Transit_times_TrES2}.
We used the ephemeris of Holman et al. (2007) to compute ''observed minus calculated'' (O-C) residuals for all 44 transit times. Fig. \ref{O_C_TrES2} shows the differences between the observed and predicted times of the center of the transit, as a function of epoch. The dashed line represents the ephemeris given by Holman et al. (2007). We found a negative trend in this (O-C)-diagram. Thus, we refine the ephemeris using the linear function for the heliocentric Julian date of midtransit (Eq. \ref{Elements}) where epoch $E$ is an integer and $T_{0}$ the midtransit to epoch 0.
\begin{equation}
\label{Elements}
T_{\mathrm{c}}=T_{0}+P\cdot E
\end{equation}
For an exact determination of the ephemeris we plotted the midtransit times over the epoch and did a linear $\chi^{2}$-fit. We got the best $\chi^{2}$ with \begin{center}
$T_{0}(\mathrm{HJD})\,=\,(2453957.63492 \pm 0.00013)\,\mathrm{d}$ \end{center} and an orbital period of \begin{center} $P\,=\,(2.470614\,\pm\,0.000001)\,\mathrm{d}$.\end{center} Our values for the $T_{0}$ and the period are different by 11.2 and 0.6\,s, respectively, compared to the ones from Holman et al. (2007). The resulting ephemeris which represents our measurements best is 
\begin{equation}
\label{ephemeris_new_TrES2}
T_{\mathrm{c}}\mathrm{(HJD)}\,=\,(2453957.63492\,+\,E\cdot 2.470614)\,d
\end{equation}

\section{Absolute photometry}

\begin{figure}
  \centering
  \includegraphics[width=0.4\textwidth]{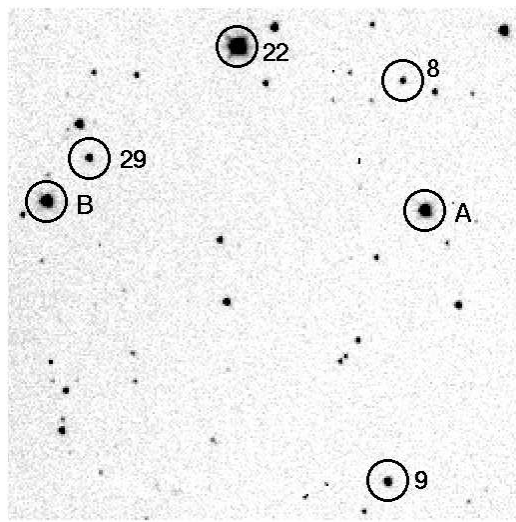}
  \caption{A 9.4$'\,\times$\,9.4$'$ 60\,s $I$-band CTK-image section of standard star field \#1 defined by Galad\'i-Enr\'iquez et al. (2000). The circles marks the used standard stars. For a definition of the labels see Table \ref{UBVRI_photometry}. The two Landold stars SA\,44\,28 and SA\,44\,113 are labelled with ''A'' and ''B'', respectively. North is up; east is to the left.}
  \label{standard_stars}
\end{figure}

To derive absolute photometric $V$, $R$ and $I$ magnitudes we observed the TrES-2 parent star on 2008 July 28, a photometric night. In order to perform the transformation to the standard system we observed standard star field \#1 defined by Galad\'i-Enr\'iquez et al. (2000) which is located around the two Landolt standard stars SA\,44\,28 and SA\,44\,113 (Landolt 1983). Altogether we took four images of the standard star field at different airmasses and a sequence of 32 TrES-2 parent star images in each filter. We measured the instrumental magnitudes of 6 standard stars - the two Landolt stars and four secondary standard stars ($UBVRI$ photometry given in Table \ref{UBVRI_photometry}, indentification chart shown in Fig. \ref{standard_stars}) - in each frame and calculated the zero point correction $c$ and the first order extinction coefficient $k$. The results for each filter are shown in Table \ref{c_and_k}. \\ During the analysis it turned out, that there is a faint object near the TrES-2 host star. We chose a larger aperture for the absolute photometry than for differential photometry (10 pixels instead of 5 pixels) to include this faint star in our measurements. With the help of the flux ratio of the two stars we could calculate the individual brightnesses. We finally derive the Bessell $V$, $R$ and $I$ magnitudes of the TrES-2 parent star obtained from the average of all 32 individual measurements. The errors given correspond to the standard deviation of these measurements:
\begin{center}
$V\,=\,11.40\,\pm\,0.02$\,mag \\ $R\,=\,11.08\,\pm\,0.02$\,mag \\ $I\,=\,10.86\,\pm\,0.05$\,mag \\
\end{center}
Our value for the $V$, $R$ and $I$ magnitudes are in good agreement with the ones published by O'Donavan et al. 2006 and Droege et al. 2006.

\begin{table}
\centering
\caption{zero point correction $c$ (normalized to 1\,s exposure time) and first order extinction coefficient $k$ for each filter}
\label{c_and_k}
\begin{tabular}{cr@{\,$\pm$\,}lr@{\,$\pm$\,}l}
\hline
Filter & \multicolumn{2}{c}{$c$} & \multicolumn{2}{c}{$k$} \\ \hline 
$V$ & 18.96 & 0.03 & 0.27 & 0.02 \\
$R$ & 18.94 & 0.02 & 0.21 & 0.02 \\
$I$ & 18.42 & 0.02 & 0.16 & 0.01 \\ \hline
\end{tabular}
\end{table}

\section{The second dip}
\label{2nd_dip}

\subsection{Detection of a second dip}

In our first observation of a transit of TrES-2 (see Fig. \ref{TrES2_Sysrem} a.) the light curve shows an interesting behavior in the last part of the observations. After the transit is completed, the flux again decreases for a short time (about 30 min). The depth of this event amounts to 0.8\,\%, which is about half the depth of the transit. To check whether the dip is a real, reproducible event and no measuring error, we scheduled TrES-2 for our regular monitoring program. An indication of the existence of the dip is published by O'Donovan et al. (2006) in a TELAST $R$-band light curve in their Fig. 1, where one can clearly recognize a brightness drop.\\ On 2007 May 3 we succeeded a second time to observe a complete transit of TrES-2 (shown in Fig \ref{TrES2_Sysrem} c.). Again the dip is clearly visible after the transit is finished.

\subsection{Further observations}
\label{dip_disappearance}

As shown in section \ref{observations} we observed ten different transits. We could detect the second dip only after our first two transit observations (Fig. \ref{TrES2_Sysrem} a. and c.). Detailed analysis of the midtime of the dip showed that the dip moved away from the center of the transit (see Table \ref{Dip}). However, this does not follow a linear relationship.
\begin{table}
\caption{Movement of the dip relative to the center of the transit}
\label{Dip}
\begin{tabular}{lcc}
\hline
Date & Distance$^{a}$ [h] & projected\\
 & & separation$^{b}$ [$\mu$as] \\ \hline 
2006 August 10 & 1.07 & 19 \\
2007 March 13 & 1.28 & 23 \\
2007 May 3 & 1.80 & 32 \\ \hline
\end{tabular}
\\$^{a}$ Time between center of the transit and center of the dip\\
$^{b}$ calculated for published distance of the TrES-2 host star (Sozzetti et al. 2007) and semi major axis of TrES-2 (O'Donovan et al. 2006)
\end{table}
\\ After using Sys-Rem for the light curves the results change.  For the observations from 2007 March 13 the dip disappeared after removing systematic effects (Fig. \ref{TrES2_Sysrem} b.) while it became stronger for the transit of 2007 May 3 (Fig. \ref{TrES2_Sysrem} d.).

\subsection{Possible explanations}

We develop some ideas which may explain the existence of the dip. Altogether, we construct four hypotheses. At the present time we cannot confirm a definitive solution. We point out that from the available data it is not possible to rule out that the second dip is photometric false alarm.

\subsubsection{A nearby variable star}

\begin{figure}[h]
  \centering
  \includegraphics[width=0.48\textwidth]{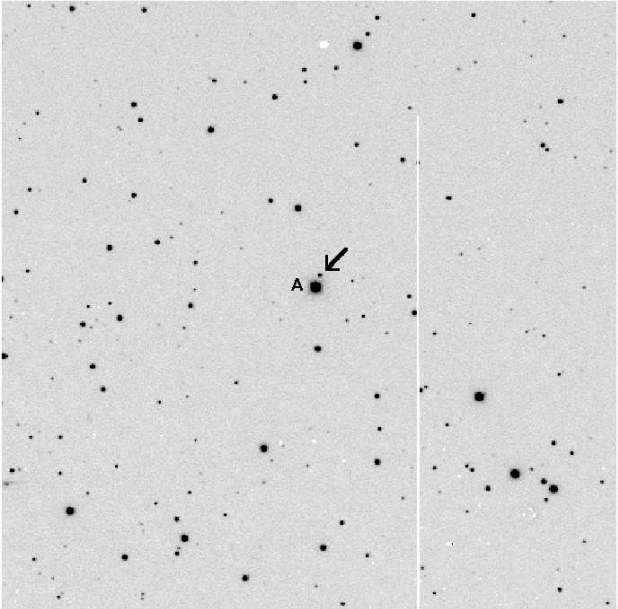}
  \caption{ A 8.5$'\,\times$\,8.5$'$ $R$-band image of the Wendelstein Observatory CCD-camera MONICA. The TrES-2 host star ia labelled with ''A''. The arrow mark the nearby star that is not resolved on images of the CTK. North is up; east is to the left.}
  \label{Position_nearby_star}
\end{figure}

\begin{figure}[h]
  \centering
  \includegraphics[width=0.48\textwidth]{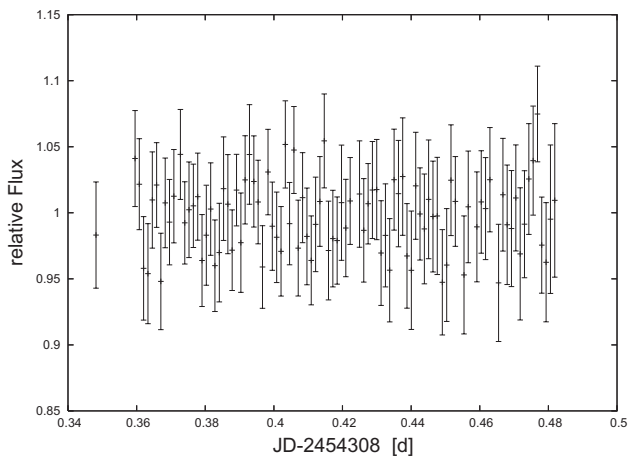}
  \caption{Wendelstein $R$-band photometry of the faint object close to the TrES-2 parent star from 2007 July 26.}
  \label{LC_nearby_star}
\end{figure}

On the CTK images we could identify another object close to the TrES-2 parent star due to elongation of the PSF. Because of the large pixel scale of the CTK this object and the TrES-2 parent star cannot be resolved independently. This faint object is within the aperture when doing the photometry. If this star is variable, it could produce a dip.\\ Again there are two explanations why the dip vanished. Because of the proper motion ($2.9\,\frac{\mathrm{mas}}{\mathrm{yr}}$ in right ascension and $-3.4\,\frac{\mathrm{mas}}{\mathrm{yr}}$ in declination, H\o{}g et al. 2000) of the TrES-2 parent star it could have moved away from the background star so that it lied no longer inside the aperture. The dip would be no longer observable. If this variable star is a blended eclipsing binary it could have an orbital period, such that in all our further observations no eclipse happened.\\ We checked this hypothesis by using a camera with better pixel scale. Obervations on the Wendelstein Observatory showed constant light for the nearby star. A $R$-band image of the Wendelstein observations and the corresponding light curve are shown in Fig. \ref{Position_nearby_star} and Fig. \ref{LC_nearby_star}. However, this result does not disprove the hypothesis of the blended eclipsing binary or another variable background star.

\subsubsection{An additional planet}

\begin{figure}[h]
  \centering
  \includegraphics[width=0.48\textwidth]{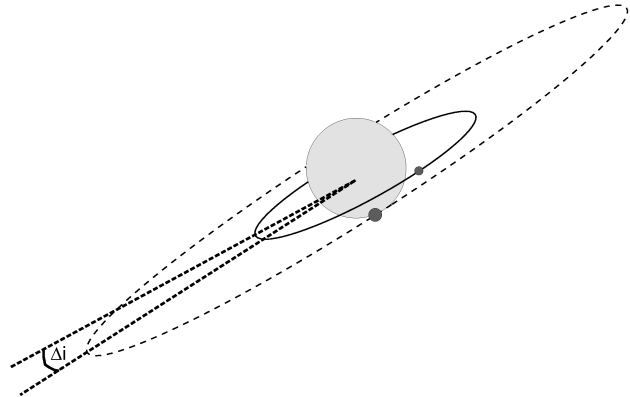}
  \caption{TrES-2 with a second outer planet whose inclination is only minimally different to the inclination of the orbit of TrES-2.}
  \label{TrES2_2nd_planet}
\end{figure}

\begin{figure}[h]
  \centering
  \includegraphics[width=0.48\textwidth]{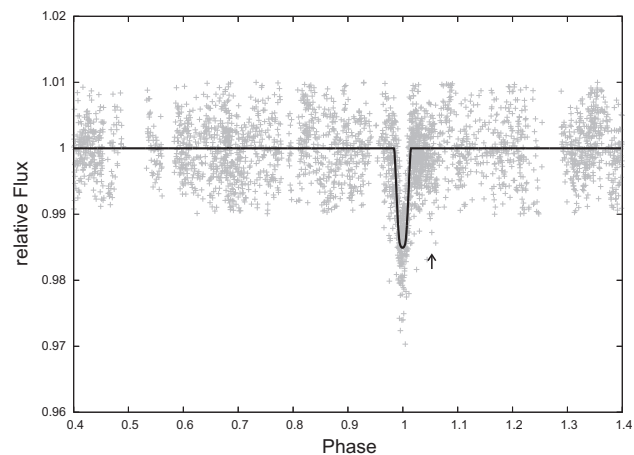}
  \caption{More than 3000 individual observations shown -  phase-folded according to the ephemeris given in eq. \ref{ephemeris_new_TrES2}. The arrow marks a possible second dip.}
  \label{TrES2_phase_diagram}
\end{figure}

Another idea is the existence of a larger outer planet. The known planet TrES-2 shows a grazing transit in front of its parent star. If the inclination of an outer planet is only minimally different from the orbit of TrES-2, the second planet could be much more grazing and could show a shorter duration. This situation is shown in Fig. \ref{TrES2_2nd_planet}: The farther out the planet the less the inclination difference, so that such a transit is observable.
\\Because the dip is always 1\,-\,2\,h after the center of the transit, both planets would have to be in mean-motion resonances. Because the outer planet needs much longer for one orbit then TrES-2, that would be an explanation for the non-detection of the dip in some observations.\\ The dip was detected after the transit events corresponding to epochs 0, 87 and 108 according to eq. \ref{ephemeris_new_TrES2}. We could not find a mean-motion resonance that fits to all three times. But if the dip on 2007 March 13 (epoch 87) is only a systematic effect, but is real in the other two light curves as shown is section \ref{dip_disappearance}, an outer planet in mean-motion resonance could again be possible. For example, the 4:1 and 9:1 resonances are consistent with all other light curves. \\ A similar situation arises if a faster inner planet shows a strong grazing transit. If the planets are in mean-motion resonance, the inner planet has to complete several orbits while TrES-2 orbits its star once.\\ Freistetter et al. (2009) show that in the region between the star and TrES-2 up to 0.03\,AU an additional planet can exist. They performed a detailed study of the stability inside the mean-motion resonances. All resonances from 0.011\,AU (6:1) to 0.026\,AU (5:3) are stable, all resonances close to the planet, from 0.028\,AU (3:2) to 0.032\,AU (5:4), are unstable. To check whether such a resonance can be found in our data, we tried to observe a complete orbital period in order to detect further dips for example in the half, one third or two thirds of the phase. In Fig. \ref{TrES2_phase_diagram} one can see the resulting phase diagram. We could not find any further dips in other parts of the phase.\\ At the moment, we could not confirm the hypothesis that a second planet in the system causes the dip.

\subsubsection{Background star}
\label{background_star}

\begin{figure}[h]
  \centering
  \includegraphics[width=0.15\textwidth]{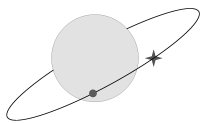}
  \caption{Origin of the dip by transiting a stationary background star.}
  \label{TrES2_background_star}
\end{figure}

As another explanation for the dip, we assume a background giant with a very low projected separation to the TrES-2 parent star. Both stars cannot be separated visually. The orbit of TrES-2 slips over the background star so that TrES-2 eclipses the background star after the actual transit is finished. Fig \ref{TrES2_background_star} illustrates this situation.
\\ However, because of the proper motion, the TrES-2 parent star has moved 3.27\,mas in the time between epochs 0 to 108. If the background star hypothesis is correct, the dip would have disappeared after a short amount of time. One can of course assume a non-stationary background star with corresponding proper motion to explain the dip.

\subsubsection{Companion}

\begin{figure}[h]
  \centering
  \includegraphics[width=0.48\textwidth]{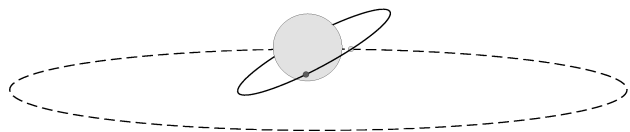}
  \caption{Origin of the dip by transiting a companion of the TrES-2 parent star (difference in inclination exaggerated)}
  \label{TrES2_companion}
\end{figure}

One of the most likely explanations for the dip at the moment is shown in Fig. \ref{TrES2_companion}.
Like in section \ref{background_star} we assume that TrES-2 transits another star along its orbit. However, it is no longer a background star, but a wide companion of the TrES-2 parent star. This companion can be very small (short transit) or the transit can be grazing.\\ The small movement of the dips of $13\,\mu as$ in 266\,d can be explained with little orbital motion of the companion. The absence of the dip in some light curves is explainable, too. Because of gravitational influence of the companion the planet's orbit precesses. Hence, sometimes TrES-2 transits the companion and sometimes not. The non-detection of the dip in any light curve after May 2007 could mean that the companion has moved so far on its orbit that TrES-2 can no longer transit it.\\ First hint for the validity of this hypotheses gave Daemgen et al. (2009). Using high-resolution ,,Lucky Imaging'' with AstraLux at the 2.2\,m Calar Alto telescope they found a companion candidate (physical association needs to be confirmed by second epoch observations) close to the TrES-2 host star. They gave a projected separation between the TrES-2 host star and the companion candidate of $(1.089\,\pm\,0.008)\,''$. Our values for the projected separation of the dip relative to the center of the transit (see Table \ref{Dip}) is five orders of magnitude lower than the projected separation of the companion candidate Daemgen et al. (2009) found. Thus, this companion candidate can not be the origin of the second dip in our light curves.

\section{Summary and Conclusions}

With the Cassegrain-Teleskop-CCD-Kamera of the University Observatory Jena we observed several transit events and almost a complete orbital period of the known exoplanet TrES-2.\\ We determined the orbital period to be slightly smaller ($\sim$ 0.6\,s) than previously expected. \\ The mean photometric precision of the $V$\,=\,11.4\,mag star is 0.007\,mag and the precision in the determination of the transit times is $\approx$\,0.0017\,d. This allows us to register transit time variations of around 150\,s. We did not find any indication of timing anomalies caused by additional planets or moons.\\ From our observations in one photometric night we were able to derive absolute photometric Bessell $V$, $R$ and $I$ magnitudes. Our estimates are in good agreement with previous published values (O'Donavan et al. 2006, Droege et al. 2006).\\ In our first observations of a transit of TrES-2 we could detect a second dip after the end of the transit. We tried to explain its existence. Four different theories have been created: a nearby variable star or a blended eclipsing binary, an additional planet in the system, a transit over a background star or a transit over a wide companion of the TrES-2 host star after the actual transit is finished. Up to now none of these theories could be rejected or confirmed as a definitive solution. Still there is also the possibility of a photometric false alarm.\\ We will continue observing TrES-2 to confirm the existence of the dip and search for transit time variations for the next few years to decades. We are also working on methods to improve the precision of our transit times. \\ The transit observations at the University Observatory Jena provide anchors for future searches for transit time variations.

\acknowledgements
The authors would like to thank Guillermo Torres for his encouraging comments to this work. Furthermore, we thank the GSH Observer Team for the nightly observations. SR and MV acknowledge support from the EU in the FP6 MC ToK project MTKD-CT-2006-042514. RN acknowledges general support from the German National Science Foundation (Deutsche Forschungsgemeinschaft, DFG) in grants NE 515/13-1, 13-2, and 23-1. AK acknowledges support from DFG in grant KR 2164/8-1. TOBS acknowledges support from Evangelisches Studienwerk e.V. Villigst. TR would like to thank the DFG for financial support (grant NE 515/23-1). TE and MMH acknowledge support from the DFG in SFB-TR 7 Gravitational Wave Astronomy. M. Moualla thanks the government of Syria for financial support. Moreover we thank the technical staff of the University Observatory Jena, especially Tobias B{\"o}hm.\\ This work has used data obtained by various observers collected in Exoplanet Transit Database, http://var.astro.cz/ETD. Some of the mid-transit times in this publication are from listings in the Amateur Exoplanet Archive.

\end{document}